\documentclass[12pt]{article}
\usepackage{epsf}
\def\be{\begin{equation}}
\def\ee{\end{equation}}
\def\bea{\begin{eqnarray}}
\def\eea{\end{eqnarray}}

\def\pp{\psi(2S)}
\def\jp{J/\psi}
\def\ypp{\Upsilon(2S)}
\def\yp{\Upsilon}

\def\p{\vec p}

\begin{document}
\begin{titlepage}
\begin{center}
{\Large \bf William I. Fine Theoretical Physics Institute \\
University of Minnesota \\}
\end{center}
\vspace{0.2in}
\begin{flushright}
FTPI-MINN-12/02 \\
UMN-TH-3028/12 \\
January 2012 \\
\end{flushright}
\vspace{0.3in}
\begin{center}
{\Large \bf Heavy quark spin symmetry breaking in near-threshold $J^{PC}=1^{--}$ quarkonium-like resonances.
\\}
\vspace{0.2in}
{\bf M.B. Voloshin  \\ }
William I. Fine Theoretical Physics Institute, University of
Minnesota,\\ Minneapolis, MN 55455, USA \\
and \\
Institute of Theoretical and Experimental Physics, Moscow, 117218, Russia
\\[0.2in]

\end{center}

\vspace{0.2in}

\begin{abstract}
A mixing of near-threshold quarkonium-like resonances with heavy meson-antimeson pairs results in an enhancement of heavy quark spin symmetry breaking, since the meson pairs are not eigenstates of the heavy quark spin. The decomposition of $P$-wave states of meson pairs in terms of the heavy-quark-pair spin states is considered in the channel with $J^{PC}=1^{--}$, which is directly produced in $e^+e^-$ annihilation. Specific processes are suggested for experimental study of the effects of the mixing with heavy meson pairs and of the internal spin structure of  bottomonium and charmonium resonances.  

\end{abstract}
\end{titlepage}

\section{Introduction}
The strength of the interaction depending on the heavy quark spin is proportional to the inverse of the heavy quark mass, so that in the limit of infinite mass the spin of the heavy quark is conserved. This well known behavior is prominently illustrated by hadronic transitions between lower states of  bottomonium and charmonium: e.g. the transitions $\ypp \to \yp(1S) \, \eta$ and $\pp \to \jp \, \eta$, requiring rotation of the spin of the heavy quark pair, are significantly suppressed relative to the corresponding heavy quark spin conserving transitions $\ypp \to \yp(1S) \, \pi \pi $ and $\pp \to \jp \, \pi \pi$~\cite{pdg}.
The observed ratio of the $\eta$ emission rate to that of the pion pair in these transitions is fully in line with the theoretical estimates~\cite{vz,mvc}. However, a completely different behavior is observed~\cite{babar} in the decays of the $\Upsilon(4S)$ resonance, where the rate of the transition $\Upsilon(4S) \to \yp(1S) \, \eta$ actually exceeds that of $\Upsilon(4S) \to \yp(1S) \, \pi \pi$. It has been pointed out~\cite{mvu} that the enhancement of the former rate can result from an admixture in the $\Upsilon(4S)$ of a four-quark state with a different alignment of the spin of the heavy quark pair (relative to the spin of the resonance), with a mixing with $B \bar B$ heavy meson pairs being a natural source of such an admixture, which clearly corresponds to an enhanced effect of the heavy quark spin symmetry breaking.

It can be argued on general grounds that the effects of the deviation from the heavy quark spin symmetry due to mixing with states of heavy meson pairs should be significantly enhanced for the quarkonium-like states in a mass band near the open flavor threshold. Indeed, in the heavy mesons the deviation at a finite heavy quark mass from the spin symmetry limit results in the splitting $\mu$ between the masses of the vector and the pseudoscalar mesons, $\mu \approx 46\,$MeV for the $B$ mesons, and $\mu \approx 140\,$MeV for the $D$ mesons. For the quarkonium-like states whose masses are well away from the thresholds for the heavy meson pairs, i.e. separated from these thresholds by a mass gap $\Delta M \gg \mu$, the possible effects of the spin-symmetry breaking in the mesons can be estimated as proportional to the dimensionless ratio $\mu/\Delta M$. Clearly, this parameter becomes of order one for the quarkonium-like resonances in the immediate vicinity of the threshold region, such as e.g. $\Upsilon(4S)$ and $\Upsilon(5S)$ for bottomonium. 

An extreme example of apparent breaking of the heavy quark spin rules is provided by the recently found $Z_b(10610)$ and $Z_b(10650)$ isovector resonances~\cite{bellez}, which decay with comparable rate to the levels of ortho-bottomonium: $Z_b \to \Upsilon(nS) \, \pi$, ($n=1,2,3$),  as well as to para-bottomonium: $Z_b \to h_b(kP) \, \pi$, ($k=1,2$). This behavior is naturally explained~\cite{bgmmv} if the $Z_b$ resonances are interpreted as `molecular' states, i.e. as threshold resonances made from $S$ wave meson-antimeson pairs with the quantum numbers $I^G(J^P)=1^+(1^+)$: $Z_b(10650) \sim B^* \bar B^*$, and $Z_b(10610) \sim (B^* \bar B - \bar B^* B)$. The heavy meson pairs in the states with quantum numbers $I^G(J^P)=1^+(1^+)$ are not  eigenstates of the total spin of the $b \bar b$ quark pair, $S_H=0^-_H$ or $S_H=1^-_H$, but rather are two orthogonal completely mixed states~\cite{bgmmv}:
\begin{eqnarray} 
&&Z_b(10610) \sim (B^* \bar B - \bar B^* B) \sim {1 \over \sqrt{2}} \, \left ( 0^-_H \otimes 1^-_{SLB} + 1^-_H \otimes 0^-_{SLB} \right )~, \nonumber \\
&&Z_b(10650) \sim B^* \bar B^*  \sim {1 \over \sqrt{2}} \, \left ( 0^-_H \otimes 1^-_{SLB} - 1^-_H \otimes 0^-_{SLB} \right )~,
\label{mix}
\end{eqnarray}
where $0^-_{SLB}$ and $1^-_{SLB}$ stand for the two possible spin states of the `rest' degrees of freedom besides the heavy quark spin. In other words, these are the two possible $J^P=1^+$ states of an $S$-wave pair of heavy mesons in the limit of spinless $b$ quark (`$SLB$' states). In this picture and due to the heavy quark spin symmetry the observed decays of the $Z_b$ resonances to $\Upsilon(nS) \, \pi$ proceed due to the presence of the ortho- ($1^-_H$) heavy quark spin state in each of the resonances, while the transitions to the para- states of bottomonium, proceed due to the part of the spin wave function with $0^-_H$. 

A complete classification of $S$-wave threshold states of heavy meson pairs in terms of their $S_H \otimes S_{SLB}$ structure is described in Refs.\cite{mvw,mp}. Of these states two more states with $J^P=0^+$ made of $B \bar B$ and $B^* \bar B^*$ also contain mixtures of ortho- and para- heavy quark pairs.

In this paper a similar analysis in terms of the spin of the heavy quark pair and the angular momentum of the `rest' degrees of freedom is applied to the states of heavy meson pairs with isospin zero and $J^{PC}=1^{--}$. This channel is of a special interest due to the direct formation of such states in $e^+e^-$ annihilation.  Clearly, these quantum numbers correspond to a $P$-wave relative motion of the mesons~\footnote{A possible presence of an $F$ wave for a $B^* \bar B^*$ pair can be neglected in the near-threshold region.}. It is necessary to emphasize that unlike the isovector states, considered~\cite{bgmmv,mvw,mp} in connection with the $Z_b$ resonances, and which are in fact states of a heavy meson pair, the isoscalar $J^{PC}=1^{--}$ states of heavy meson pairs should be considered as an admixture to the pure heavy quarkonium states, of which the ones produced in $e^+e^-$ annihilation are $^3 \! S_1$ states of the heavy quark pair. In the considered here classification  in terms of their $S_H \otimes S_{SLB}$ structure, the quarkonium $^3 \! S_1$ states are $1^-_H \otimes 0^+_{SLB}$, since the (absent) `rest' degrees of freedom are in the vacuum state corresponding to $0^+_{SLB}$. A possible small admixture of $^3 \! D_1$ heavy quark pair, which is to be classified as that of a $1^-_H \otimes 2^+_{SLB}$ arises in the second order in the breaking of the heavy quark symmetry and is neglected here. 

In what follows, for definiteness and simplicity of the notation, the properties of the bottomonium-like states and of $B^{(*)}$ meson-antimeson pairs are discussed. An application to similar properties of charmonium and $D^{(*)}$ mesons will be mentioned separately.

The rest of the paper is organized as follows. In Sec.~2 the transformation from the states of meson pairs to the eigenstates of the heavy quark spin is derived. In Sec.~3 an application of the spin symmetry to production of heavy meson pairs in $e^+e^-$ annihilation is discussed, and in Sec.~4 properties of specific bottomonium-like and charmonium-like vector resonances are considered. Finally, the discussion and results are summarized in Sec.~5. 

\section{Spin structure of the $J^{PC}=1^{--}$ heavy meson pairs }

There are four different $P$-wave states of the heavy mesons with $J^{PC}=1^{--}$:
\bea
B \bar B&:&~~~~ p_i \, (B^\dagger B)~; \nonumber \\
{B^* \bar B - \bar B^* B \over \sqrt{2}}&:& ~~~{i \over 2} \, \epsilon_{ijk} p_j \, (B^{*\dagger}_k B - B^*_k B^\dagger)~; \nonumber \\
(B^* \bar B^*)_{S=0}&: &~~~{p_i \over \sqrt{3}} \, (B^{*\dagger}_j B^*_j)~; \nonumber \\
(B^* \bar B^*)_{S=2}&:& ~~~ \sqrt{ 3 \over 5} \, {p_k \over 2} \,  \left ( B^{*\dagger}_i B^*_k + B^{{*\dagger}}_k B^*_i - {2 \over 3} \, \delta_{ik} \,  B^{{*\dagger}}_j B^*_j \right )~.
\label{b4}
\eea
The states ($B^* \bar B^*)_{S=0}$ and $(B^* \bar B^*)_{S=2}$  correspond to two possible values of the total spin $S$ of the $B^* \bar B^*$ meson pair.
The wave functions in the r.h.s are written in terms of the c.m. momentum $\vec p$ and the wave functions of the pseudoscalar and vector mesons and have the same normalization for each state. 

The four states of the meson pairs in Eq.(\ref{b4}) are not eigenstates of either the operator of the total spin $\vec S_H$ of the heavy quark pair, nor of the operator $\vec J_{SLB}= \vec S_{SLB}+ \vec L$, describing the angular momentum in the limit of spinless $b$ quark. Clearly, there are four possible combinations of such eigenstates that match the overall quantum numbers $J^{PC}=1^{--}$: 
\be
\psi_{10}=1^{--}_H \otimes 0^{++}_{SLB}\,, ~~~\psi_{11}=1^{--}_H \otimes 1^{++}_{SLB}\,,~~~ \psi_{12}=1^{--}_H \otimes 2^{++}_{SLB}\,, ~~{\rm and}~~ \psi_{01}=0^{-+}_H \otimes 1^{+-}_{SLB}\,.
\label{psis}
\ee
The first three of these combinations involve an ortho- state of the $b \bar b$ pair with different alignment of the total spin $S_H=1$ relative to the total angular momentum of the state, while the fourth combination involves a para-  $b \bar b$ state, i.e. with $S_H=0$, while the overall angular momentum is provided by that of the `rest' degrees of freedom, $J_{SLB}=1$ (and a negative $C$ parity, which in simple terms of `the light quark pair' $q \bar q$ corresponds to a $^1\!P_1$ state). 

The explicit expansion of the four states in Eq.(\ref{b4}) in terms of the four eigenfunctions $\psi_{ab}$ can be readily found, similarly to the method used in Ref.~\cite{bgmmv} by replacing in Eq.(\ref{b4}) the wave functions of the $B^{(*)}$ mesons with interpolating expressions in terms of nonrelativistic spinors $b$ ($b^\dagger$) for the $b$ (anti)quark and the nonrelativistic spinors $q$ and $q^\dagger$ for the `rest' degrees of freedom in the mesons, 
$B \sim (b^\dagger q)$, $B^*_i \sim (b^\dagger \, \sigma_i \, q)$, and performing the Fierz transformation, e.g.
$$ (b^\dagger q) (q^\dagger b)= - {1 \over 2} \, (b^\dagger \, \sigma_i \, b) (q^\dagger \, \sigma_i \, q) - {1 \over 2} \, (b^\dagger  b) (q^\dagger  q)~.$$
The result has the form:
\bea
B \bar B &:& ~~~{1 \over 2  \sqrt{3}} \, \psi_{10}+ {1 \over 2} \, \psi_{11}+{\sqrt{5} \over 2 \sqrt{3}} \, \psi_{12} + {1 \over 2} \, \psi_{01}~; \nonumber \\
{B^* \bar B - \bar B^* B \over \sqrt{2}}&:&  ~~~{1 \over  \sqrt{3}} \, \psi_{10}+ {1 \over 2} \, \psi_{11}-{\sqrt{5} \over 2 \sqrt{3}} \, \psi_{12}~; \nonumber \\
(B^* \bar B^*)_{S=0}&:& ~~~ - {1 \over 6} \, \psi_{10}- {1 \over 2 \sqrt{3}} \, \psi_{11}-{\sqrt{5} \over 6} \, \psi_{12} + {\sqrt{3} \over 2} \, \psi_{01}~; \nonumber \\
(B^* \bar B^*)_{S=2}&:& ~~~{\sqrt{5} \over  3} \, \psi_{10} - {\sqrt{5} \over 2 \sqrt{3}} \, \psi_{11} + {1 \over 6} \, \psi_{12}~.
\label{bp4}
\eea
One can easily check that the matrix of the transformation from the $H \otimes SLB$ eigenstates to the states of the meson pairs is orthogonal.

\section{Production of heavy meson pairs in $e^+e^-$ annihilation}
The heavy mesons are produced by the electromagnetic current of the heavy quark, e.g. $(\bar b \, \gamma_\mu b)$, which in the nonrelativistic near-threshold region corresponds to the structure $1^{--}_H \otimes 0^{++}_{SLB}$. Therefore in the limit of exact heavy quark spin conservation the relative amplitudes for production of the four states of the meson pairs are given by the coefficients of $\psi_{10}$ in Eq.(\ref{bp4}):
\bea
&&A(e^+e^- \to B \bar B): A( e^+e^- \to B^* \bar B +c.c.):A \left [ e^+e^- \to (B^* \bar B^*)_{S=0} \right ]: A \left [ e^+e^- \to (B^* \bar B^*)_{S=2} \right ] \nonumber \\
&&= {1 \over 2  \sqrt{3}}: {1 \over  \sqrt{3}} : - {1 \over 6}: {\sqrt{5} \over  3}~.
\label{ra}
\eea
These ratios  give rise to the relation between the production cross section $\sigma$ for each channel, normalized to the corresponding $P$-wave phase space factor $v^3$ with $v$ being the c.m. velocity~\footnote{ It should be noted that taking into account the difference of the phase space factors is beyond the exact symmetry limit and is not necessarily justified}:
\bea
&&{\sigma \over v^3}(e^+e^- \to B \bar B): {\sigma \over v^3}( e^+e^- \to B^* \bar B +c.c.):{\sigma \over v^3} \left [ e^+e^- \to (B^* \bar B^*)_{S=0} \right ]: {\sigma \over v^3} \left [ e^+e^- \to (B^* \bar B^*)_{S=2} \right ] \nonumber \\
&&= 1: 4 : {1 \over 3} :{20 \over 3}~.
\label{rs4}
\eea
If the states of the vector meson pairs with $S=0$ and $S=2$ are not resolved and only the total yield of the mesons is measured, as is the case in the existing data, the ratio for the factors $\sigma/v^3$ is given by
\be
{\sigma \over v^3}(e^+e^- \to B \bar B): {\sigma \over v^3}( e^+e^- \to B^* \bar B +c.c.):{\sigma \over v^3} ( e^+e^- \to B^* \bar B^*)=1:4:7~.
\label{rs3}
\ee
This relation, pointed out long ago~\cite{drgg}, is a direct consequence of exact heavy quark spin symmetry. In fact this relation is in a dramatic contradiction with most of the known data on production of $B^{(*)}$ mesons as well as of the charmed mesons~\cite{cleoc,bellec,babarc} in $e^+e^-$ annihilation near the corresponding thresholds, and this contradiction is a good illustration of the strong breaking of the spin symmetry in near-threshold region. Such behavior is known already from the early observations of the charmonium-like peak $\psi(4040)$ where the yield of the vector meson pairs $D^* \bar D^*$ greatly exceeds that implied by the relation (\ref{rs3}). 

It appears reasonable to consider the relative yield of different heavy meson pairs as an indicator of the internal structure of a quarkonium-like state in terms of the $H \otimes SLB$ decomposition described by the relations (\ref{bp4}). In this regard it would be helpful if the yield of two possible waves for the vector meson pairs were resolved experimentally, since these two waves correspond to substantially different spin-symmetry structures. In particular the $S=0$ wave has a nonvanishing projection on the eigenstate $\psi_{01}$ corresponding to para- state of the heavy quark pair, while for the $S=2$ wave the admixture of the para- state is zero. One way of resolving the contribution of the two waves is by the angular distribution in the angle $\theta$ between the direction of the vector meson pair and that of the $e^+e^-$ beams:
\bea
&&{d \over d \cos \theta} \, \sigma \left [ e^+e^- \to (B^* \bar B^*)_{S=0} \right ]\, \propto  \, 1-\cos^2 \theta~; \nonumber \\
&&{d \over d \cos \theta} \, \sigma \left [ e^+e^- \to (B^* \bar B^*)_{S=2} \right ] \, \propto \, 1- {1 \over 7} \, \cos^2 \theta~.
\label{vv2}
\eea

\section{Specific quarkonium-like resonances}
\subsection{$\Upsilon(4S)$}
Due to the proximity of $\Upsilon(4S)$ to the $B \bar B$ threshold it is likely  that the four-quark admixture in it is dominated by the pseudoscalar meson pair. The $H \otimes SLB$ structure of such admixture is given by the first line in Eq.(\ref{bp4}). One can notice that in addition to the state $\psi_{11}$ that is presumably~\cite{mvu} responsible for the greatly enhanced rate of $\Upsilon(4S) \to \Upsilon(1S) \, \eta$ the $B \bar B$ component contains also an admixture of the state $\psi_{01}$ with a spin-singlet $b \bar b$ pair. Therefore if the assumption of the dominance of a $B \bar B$ admixture in $\Upsilon(4S)$ is correct, one should expect an enhancement of the (yet unknown) decay $\Upsilon(4S) \to \eta_b \, \omega$, which is very strongly suppressed for a pure $b \bar b$ bottomonium by the spin symmetry. 

Furthermore, the presence of the states $\psi_{11}$ and $\psi_{12}$ should generally give rise to a dipion $D$-wave coupled to the spins of the bottomonium resonances in the transition $\Upsilon(4S) \to \yp(1S) \, \pi \pi$:
\be
A[\Upsilon(4S) \to \yp(1S) \, \pi(p_1) \pi(p_2)] = A_S \, (\vec \Upsilon^{(4)} \cdot \vec \Upsilon^{(1)})+A_D \, \left [ p_{1i} p_{2j} + p_{1j} p_{2i} - {2 \over 3} \delta_{ij} \, (\vec p_1 \cdot \vec p_2) \right ] \,  \Upsilon^{(4)}_i \Upsilon^{(1)}_j~,
\label{dwu}
\ee
where the coefficients $A_S$ and $A_D$ for respectively the spin-decoupled amplitude and the spin-coupled one are generally scalar functions of the pion momenta. In the soft-pion limit $A_D$ is a constant, while $A_S$ should be bilinear in the pion momenta and energies: $A_S = a \, \varepsilon_1 \varepsilon_2 + b (\vec p_1 \cdot \vec p_2)$~\cite{bc,mv75}~\footnote{The discussed here coefficients are related, up to a common overal normalization, to the coefficents $A$, $B$ and $C$ introduced in Ref.~\cite{bc} as: $a=B-A$, $b = A+ 2C/3$, $A_D=C$}.
The admixture of the spin-dependent $D$-wave can be tested experimentally by a deviation of the angular distribution of the leptons emerging in the decay $\yp(1S) \to \ell^+ \ell^-$ from the standard $1+ \cos^2 \theta$. Another probe can be a measurement of a correlation between the direction of a single pion momentum (e.g. $\pi^+$ in the transition $\Upsilon(4S) \to \yp(1S) \, \pi^+ \pi^-$) with the direction of the initial beams. In the absence of a spin-coupled $D$ wave the distribution is isotropic, while for a pure $D$-wave the distribution is $(1-{21 \over 47} \, \cos^2 \theta)$. In reality both amplitudes are likely to be present with the spin independent term $A_S$ being dominant, and the spin dependent part representing a `new' effect. In this case it may require a study of the full angular information to untangle the spin structure of the amplitude.
The expression for the full angular correlations in this process is given in the Appendix. 
\subsection{$\Upsilon(5S)$}
The resonance $\Upsilon(5S)$ with mass of approximately 10.87\,GeV is well above the thresholds for nonstrange $B^{(*)}$ mesons, so that the spin symmetry breaking due to the mixing with non-strange mesons should not be large. The resonance however is very close to the threshold at 10.83\,GeV for the pair of vector strange $B_s^*$ mesons, and it is likely that the mixing with such pairs is most important. In other words, it is plausible that the dominant structure of the $\Upsilon(5S)$ is a mixture of a pure-bottomonium-like state $1^{--}_H \otimes 0^{++}$ and a pair $B_s^* \bar B_s^*$. Assuming that a conversion of the latter admixture with hidden strangeness into pairs of nonstrange $B$ mesons is somewhat OZI suppressed, one can expect that the relative yield of the nonstrange mesons should be close to the symmetry limit relation (\ref{rs3}). The available data~\cite{pdg} give the ratio of the cross sections
\be
{\sigma }\left[ \Upsilon(5S) \to B \bar B \right ] : {\sigma  }\left[ \Upsilon(5S) \to B^* \bar B +c.c. \right ] :{\sigma  }\left[ \Upsilon(5S) \to B^* \bar B^* \right ]  \approx 1\, : \, 2.5 \, : \,7 ~,
\label{ry5}
\ee
which is not very far away from the symmetry relation (without an account for the difference of the phase space factors), and in fact is the only known instance, where the ratio looks in any sense like that following from the heavy quark spin symmetry. In any case, the relative yield of the non-strange mesons is significantly closer to the relation (\ref{rs3}) than that for the strange $B_s^{(*)}$ meson pairs, where the double-vector state absolutely dominates in the ratio.  The latter behavior clearly agrees with the assumption that the dominant spin symmetry breaking component in $\Upsilon(5S)$ is $B_s^* \bar B_s^*$. 

Another indication of a suppressed spin symmetry breaking in $\yp(5S)$ in the sector without hidden strangeness is provided by its dipion transitions to lower bottomonium levels. 
The resonance $\Upsilon(5S)$ is known~\cite{belleh} to experience dipion transitions to both orthobottomonium, $\Upsilon(5S) \to \Upsilon(nS) \, \pi \pi$, and parabottomonium, $\Upsilon(5S) \to h_b(kP) \, \pi \pi$. There is however an important difference between these processes. Namely, the former transitions proceed through both the $Z_b$ resonances and through a non-resonant mechanism, while in the latter transitions to the parabottomonium levels $h_b$ the non-resonant background is small (in fact compatible with zero in the available data~\cite{bellez}). The strong suppression of direct transitions to the para- states indicates that there is essentially no  presence of the state $0^{-+}_H \otimes 1^{+-}_{SLB}$ without hidden strangeness in the spin structure of the $\Upsilon(5S)$, in agreement with unbroken spin symmetry. Additionally a suppression of this state within $\yp(5S)$ is indicated by the data~\cite{bellez} on the ratio of the coupling the resonance to the channels $Z_b(10610) \, \pi$ and $Z_b(10650) \, \pi$, which data are compatible with the ratio equal to one, as it should be, given the structure (\ref{mix}) of the $Z_b$ resonances~\footnote{The latter argument is due to A.~Bondar.}.

The sector with no hidden strangeness can be further studied in the dipion transitions to the lower $\yp(nS)$ states, namely by a search for a spin-coupled dipion $D$ wave as was discussed for the $\yp(4S)$ resonance. In fact it has been noted in Refs.~\cite{bellez} that the quality of a fit to the Dalitz plot distribution for the process $\yp(5S) \to \yp(1S) \, \pi^+ \pi^-$ is significantly improved by inclusion of the $f_2(1270)$ resonance in the dipion channel. If confirmed, the contribution of this resonance would imply a presence of the spin-coupled $D$ wave, and may be interpreted as arising from an admixture in $\yp(5S)$ of $\psi_{11}$ and/or $\psi_{12}$ states without hidden strangeness. Such behavior however would be somewhat problematic to make compatible with the relation (\ref{ry5}) and the presented arguments in favor of a naturally small breaking of the heavy quark spin in $\yp(5S)$ due to mixing with pairs of nonstrange $B^{(*)}$ mesons. Possibly, some understanding of the spin and (hidden) flavor structure of the light-meson pairs with invariant mass above 1\,GeV emitted in the $\yp(5S) \to \yp(1S)$ transitions can be gained from a comparison of the $\yp(5S) \to \yp(1S) \, \pi^+ \pi^-$ and $\yp(5S) \to \yp(1S) \, K \bar K$ data.

An additional option for such study is provided by the possible dipion transitions from $\yp(5S)$ to the $1^3\!D_J$ states of bottomonium. A hint (at about $2.5 \, \sigma$ significance) at an existence of such transitions at a level feasible for observation is contained in Ref.~\cite{belleh}.  Dipion transitions between the $\yp$ resonances and the $^3\!D_J$ states of bottomonium were discussed~\cite{blmn,moxhay,kr,mvd} in terms of the multipole expansion in QCD, which does not appear to be applicable to the process $\yp(5S) \to 1^3\!D_J \, \pi \pi$. However it can be argued without relying on the multipole expansion that the form of the amplitude for this process is uniquely determined in the heavy quark symmetry limit, if one also uses the soft pion properties. Indeed, if the initial $\yp(5S)$ state is a pure $1^{--}_H \otimes 0^{++}_{SLB}$, the spin polarization of the $b \bar b$ pair coincides with the polarization vector amplitude of the resonance $\vec \yp^{(5)}$. The final $^3\!D$ state is a pure bottomonium, and its wave function factorizes into a product of the $S=1$ spin polarization vector $\vec \chi$ and the $D$ wave coordinate wave function $\phi_{ij}$. The heavy quark spin conservation implies that the amplitude for  the transitions $\yp(5S) \to ^3\!D_J \, \pi \pi$ is proportional to the scalar product $(\vec \yp^{(5)} \cdot \vec \chi)$, while the orbital $L=2$ state is generated by the pions, which thus have to be in the $D$-wave. Given that the soft pion theorems require the amplitude to vanish when the four-momentum of either of the pions goes to zero, the form of the amplitude for the coupling of the pions to the coordinate part of the $D$ state wave function is uniquely determined as $p_{1i} p_{2j} \phi_{ij}$ with $\vec p_1$ and $\vec p_2$ being the momenta of the pions. Thus the amplitudes for the transitions to all three $^3\!D_J$  states with $J=1,2,3$ are determined by one constant $A$:
\be
A \left[ \yp(5S) \to ^3\!D_J \, \pi(p_1) \pi(p_2) \right ] = A \, p_{1i} p_{2j} \yp^{(5)}_k \, P^{(J)} \chi_k \phi_{ij}~,
\label{ad}
\ee
where $P^{(J)}$ is the projector of the direct product $\chi_k \phi_{ij}$ on the state with definite $J$. 

The amplitude in Eq.(\ref{ad}) implies that the transition rate for each $J$ is proportional to the statistical weight $2J+1$, and that the distribution in the angle $\vartheta$ between the two pion momenta is the same for all three states:
\be
{d \Gamma \over d \cos \vartheta} \propto 1+ {1 \over 3} \, \cos^2 \vartheta~.
\label{vt}
\ee
A difference between the final states with different $J$ arises in the correlation of the directions of the pion momenta with the initial $e^+ e^-$ beams. The corresponding formulas for these angular correlations are given in the Appendix.

Regarding the discussed four-quark component of the $\yp(5S)$ resonance with hidden strangeness, the
composition of the   $B_s^* \bar B_s^*$ pairs in terms of the states with total spin of the meson pair $S=0$ and $S=2$ is yet unknown. If the $S=0$ state is present, then, according to Eq.(\ref{bp4}) there should be a $0^{-+} \otimes 1^{+-}_{SLB}$ component with hidden strangeness in the structure of the resonance. Such component should likely give rise to the processes $\yp(5S) \to h_b(kP) \, \eta$, $\yp(5S) \to h_b(1P) \, \eta'$ and $\yp(5S) \to \eta_b \, \phi$. Generally, an admixture of $B_s^* \bar B_s^*$ states with either total spin of the mesons introduces also $1^{--}_H \otimes 1^{++}_{SLB}$ and $1^{--}_H \otimes 2^{++}_{SLB}$ components with hidden strangeness in the spin structure of the resonance. The former should result in enhanced transitions $\yp(5S) \to \yp(1S) \, \eta$, $\yp(5S) \to \yp(1S) \, \eta'$ and $\yp(5S) \to \yp(2S) \, \eta$ as well as in a presence of the spin-coupled  $D$ wave of light mesons in the transitions $\yp(5S) \to \yp(1S) \, K \bar K$ and $\yp(5S) \to \yp(1S) \, \eta \eta$.

One of the tests of the light-flavor composition of the spin symmetry breaking components is provided by the relative yield of $\eta$ and $\eta'$. For a pure $s \bar s$ admixture one expects, e.g.
\be
{\Gamma \left [\yp(5S) \to \yp(1S) \, \eta' \right ] \over \Gamma \left [\yp(5S) \to \yp(1S) \, \eta \right ]} \approx {p_{\eta'}^3 \over 2 \, p_{\eta}^3}~~~{\rm and}~~~ {\Gamma \left [\yp(5S) \to h_b(1P) \, \eta' \right ] \over \Gamma \left [\yp(5S) \to h_b(1P) \, \eta \right ]} \approx {p_{\eta'} \over 2 \, p_{\eta}}~.
\label{etar}
\ee

\subsection{Charmonium-like vector resonances}
The scale of the heavy quark spin symmetry breaking in the charmed $D$ mesons, $\mu \approx 140\,$MeV, is larger than in the $B$ mesons due to smaller mass of the charmed quark. Therefore the discussed effects of the spin symmetry violation near open heavy flavor threshold cover a broader range of masses of charmonium-like states. In particular, the $e^+e^-$ cross section for production of charmed meson pairs with and without hidden strangeness displays an intricate behavior~\cite{cleoc,bellec,babarc} in this region of c.m. energy, that generally is quite different from the symmetry prediction in Eq.(\ref{rs3}).

As already mentioned, the resonance $\psi(4040)$ with mass just above the $D^* \bar D^*$ threshold couples most strongly to the vector meson pairs to the extent that it has been suggested~\cite{drgg} that this resonance is mostly a `molecular' state made from vector meson pairs. In line with the previous discussion it is possible that a (presumably) large four-quark component in $\psi(4040)$ can give rise to an enhanced process $\psi(4040) \to \jp \, \eta$. The transition $\psi(4040) \to \eta_c \, \omega$ may also be enhanced to a detectable level, if there is a presence of $(D^* \bar D^*)_{S=0}$ pairs in the $\psi(4040)$ internal structure.

Similar expectations also apply to the resonance $\psi(4160)$, i.e. a possible existence of the transitions $\psi(4160) \to \jp \, \eta$ and/or $\psi(4160) \to \eta_c \, \omega$ depending on the internal structure of $\psi(4160)$ in terms of $H \otimes SLC$ states. The additional peculiarity of this state is that its dipion transition to the paracharmonium level $h_c$, $\psi(4160) \to h_c \, \pi \pi$, has been observed~\cite{cleoh} at a rate comparable to  that of $\psi(4160) \to \jp \, \pi \pi$. At first glance this resembles the bottomonium processes $\yp(5S) \to h_b(kP) \, \pi \pi$. The bottomonium transitions,  proceed through the isovector  $Z_b$ resonances with essentially absent non-resonant background~\cite{bellez}. Therefore one might suggest that similar isovector resonances, $Z_c$, do exist for charmonium and that those enhance the spin symmetry breaking decay   $\psi(4160) \to h_c \, \pi \pi$. It should be noted however that the exotic resonance mechanism is possible but not absolutely necessary for explaining this process, since this can proceed due to a paracharmonium component within the $\psi(4160)$ induced by the enhanced spin symmetry breaking. Naturally, it is possible that both mechanisms contribute to the decay. In this regard a study of the transition $\psi(4160) \to h_c \, \eta$ is of great interest, since this spin symmetry breaking process is not contributed by the hypothetical $Z_c$ resonances. A hint at an observation of this decay at a $3 \sigma$ level has been reported~\cite{cleoh}. If confirmed, the $\eta$ transition would imply a presence in $\psi(4160)$ of a four-quark component with spin-singlet $c \bar c$ pair. If the light quark pair in this component is not exclusively $s \bar s$, i.e. with hidden strangeness, then the $\pi \pi$ transition can proceed through a non-resonant mechanism. In case the light quark pair is dominated by $s \bar s$, the mechanisms for the $\eta$ and $\pi \pi$ transitions to the $h_c$ state would have to be different (modulo a violation of the OZI rule). The (hidden) light flavor composition of the four-quark component can in principle be tested by measuring the ratio of the rates of $\eta$ and $\eta'$ emission: $\Gamma[\psi(4160) \to \jp \, \eta']/\Gamma[\psi(4160) \to \jp \, \eta]$.

For the still higher in mass charmonium resonance $Y(4260)$ the decays to charmed meson pairs do not appear to be significant~\cite{cleoc}, so that  breaking of spin symmetry due to mixing with meson pairs does not look plausible. A certain increase in the yield of three body final states, $D^{(*)} \bar D^{(*)} \pi$ has been observed at the energy of $Y(4260)$, however the significance of these states is not clear. The only reliably established decay channels for this resonance: $\jp \, \pi \pi$ and $\jp \, K \bar K$, do not indicate any spin symmetry breaking. However there is a hint~\cite{cleoh} at existence of the decay $Y(4260) \to h_c \, \pi \pi$. If confirmed, this process would imply a breaking of the spin symmetry, which can arise either indeed through exotic resonances $Z_c$, or require some more complicated internal structure of the state $Y(4260)$.

\section{Summary}
The breaking of the heavy quark spin symmetry is expected to be enhanced in  quarkonium-like states in the mass range near the thresholds for pairs of heavy mesons. This behavior conspicuously shows in the  $Z_b$ resonances but is also apparently present in the decay properties of vector charmonium-like and bottomonium-like resonances that can be directly produced and studied in $e^+e^-$ annihilation. The discussed here enhanced violation of the spin symmetry is attributed to mixing of the quarkonium-like states with pairs of heavy mesons. The meson-antimeson pairs are not eigenstates of the heavy quark spin operator, but rather mixtures of such states, as described, in the $J^{PC}=1^{--}$ channel, by the formulas (\ref{bp4}). The four-quark admixture in quarkonium-like resonances can thus be studied in terms of mixing with meson pairs through decay processes violating the heavy quark spin symmetry. Such processes include those requiring rotation of the total spin of the heavy quark pair, e.g. $\Upsilon(4S) \to \yp(1S) \, \eta$, $\Upsilon(5S) \to \yp(nS) \, \eta$, $\psi(4160) \to \jp \, \eta$ and a presence of spin-coupled $D$-wave in di-meson transitions of the type $\Upsilon(4S) \to \yp(nS) \, \pi \pi$, or $\yp(5S) \to \yp(1S) \, K \bar K$. Testing a presence of such $D$-wave would require an experimental study of angular correlations described by Eq.(\ref{acorr}). An even more illustrative of the spin symmetry breaking would be an observation of transitions to spin-singlet levels of quarkonium such as $\Upsilon(5S) \to \eta_b \, \omega$, $\psi(4160) \to \eta_c \, \omega$, and $\Upsilon(5S) \to h_b \, \eta$ (and, possibly, $\Upsilon(5S) \to h_b \, \eta'$), $\psi(4160) \to h_c \, \eta$. Furthermore, it has been assumed throughout the present discussion that some degree of separation between four-quark states with and without hidden strangeness is provided by the OZI rule. It is not yet known to what extent this rule can be applied within the quarkonium-like resonances, neither its dependence on the quantum numbers of the $SLB$ states is known. Therefore it would be quite instructive to study processes with different light flavor content, e.g. $K \bar K$ emission vs. $\pi \pi$, or $\eta'$ vs. $\eta$. It can be also noted that a study of internal spin structure of  $J^{PC}=1^{--}$ quarkonium-like resonances would greatly benefit from a separate measurement of the yield of vector meson pairs in the two possible $P$-wave states: one with the total spin $S$ of the meson pair equal to zero, and the other with $S=2$. These two states can be separated e.g. by the angular distribution relative to the direction of the $e^+e^-$ beams. These two states significantly differ in terms of the spin structure of the heavy quark pair, and their admixture in quarkonium-like resonances generally leads to different decay properties.

\section*{Acknowledgments}
I am grateful to Alexander Bondar for many enlightening discussions. This work is supported, in part, by the DOE grant DE-FG02-94ER40823.

\section*{Appendix}
\subsection*{Angular correlations in the decay $\yp(4S) \to \yp(1S) \, \pi \pi$.}
Introducing unit vectors $\vec n_1$ and $\vec n_2$ along respectively the direction of the initial $e^+e^-$ beams and the direction of the lepton pair in the decay of the final $\Upsilon(1S)$ resonance, the angular distribution can be found from the amplitude in Eq.(\ref{dwu}) in the form:
\bea
&&\left | A[\Upsilon(4S) \to \yp(1S) \, \pi(p_1) \pi(p_2)] \right |^2 = |A_S|^2 \, \left [ 1+ (\vec n_1 \cdot \vec n_2)^2 \right ]  - \nonumber \\
&&2 \, {\rm Re}(A_S A_D^*) \, \left [ 2 \, (\vec p_1 \cdot \vec n_1) (\vec p_2 \cdot \vec n_1)+ 2 \, (\vec p_1 \cdot \vec n_2) (\vec p_2 \cdot \vec n_2) - {4 \over 3} \, (\vec p_1 \cdot \vec p_2) -  \right . \nonumber \\
&&\left .  (\vec p_1 \cdot \vec n_2) (\vec p_2 \cdot \vec n_1) (\vec n_1 \cdot \vec n_2) - (\vec p_1 \cdot \vec n_1) (\vec p_2 \cdot \vec n_2) (\vec n_1 \cdot \vec n_2) + {2 \over 3} \, (\vec p_1 \cdot \vec p_2) (\vec n_1 \cdot \vec n_2)^2  \right ] + \nonumber \\
&&|A_D|^2 \left [ 2 \, p_1^2 p_2^2 - {2 \over 9} \, (\vec p_1 \cdot \vec p_2)^2 - p_1^2 \, (\vec p_2 \cdot \vec n_1)^2 - p_2^2 \, (\vec p_1 \cdot \vec n_1)^2 
- p_1^2 \, (\vec p_2 \cdot \vec n_2)^2 - p_2^2 \, (\vec p_1 \cdot \vec n_2)^2 - \right . \nonumber \\ 
&&\left . {2 \over 3} \, (\vec p_1 \cdot \vec n_1) (\vec p_2 \cdot \vec n_1)(\vec p_1 \cdot \vec p_2) - {2 \over 3} \, (\vec p_1 \cdot \vec n_2) (\vec p_2 \cdot \vec n_2)(\vec p_1 \cdot \vec p_2)+ \right. \nonumber \\ 
&&\left . (\vec p_1 \cdot \vec n_1)^2 (\vec p_2 \cdot \vec n_2)^2+ (\vec p_1 \cdot \vec n_2)^2 (\vec p_2 \cdot \vec n_1)^2+ {4 \over 9} \, (\vec p_1 \cdot \vec p_2)^2 (\vec n_1 \cdot \vec n_2)^2 + \right . \nonumber \\ 
&&\left . 2 \, (\vec p_1 \cdot \vec n_1)(\vec p_2 \cdot \vec n_1) (\vec p_1 \cdot \vec n_2)(\vec p_2 \cdot \vec n_2) - {2 \over 3} \, (\vec p_1 \cdot \vec p_2)(\vec n_1 \cdot \vec n_2) (\vec p_1 \cdot \vec n_1)(\vec p_2 \cdot \vec n_2)
 - \right . \nonumber \\ 
&&\left .  {2 \over 3} \, (\vec p_1 \cdot \vec p_2)(\vec n_1 \cdot \vec n_2) (\vec p_1 \cdot \vec n_2)(\vec p_2 \cdot \vec n_1) \right]~,
\label{acorr}
\eea
where $p^2$ stands for $\vec p~^2$.

\subsection*{Angular correlations in the decay $\yp(5S) \to ^3\!D_J \, \pi \pi$.}
Introducing a unit vector $\vec n$ in the direction of the beams one can find from the amplitude in Eq.(\ref{ad}) the triple correlations between $\vec n$ and the pion momenta for each final state:
\bea
&&  \sum_{\rm pol}\,\left | A[e^+e^- \to \Upsilon(5S) \to ^3\!D_1 \, \pi \pi] \right |^2= \nonumber \\
&&{3  C \over 5} \, \left [ 2 p_1^2 p_2^2 + {2 \over 9} \, (\p_1 \cdot \p_2)^2 - (\p_1 \cdot \vec n)^2 p_2^2 - (\p_2 \cdot \vec n)^2 p_1^2 + {2 \over 3} \, (\p_1 \cdot \p_2) (\p_1 \cdot \vec n) (\p_2 \cdot \vec n) \right ]~; \nonumber \\
&& \sum_{\rm pol}\,\left | A[e^+e^- \to \Upsilon(5S) \to ^3\!D_2 \, \pi \pi] \right |^2= \nonumber \\
 &&{2  C \over 3} \,\left [  p_1^2 p_2^2 +  (\p_1 \cdot \p_2)^2 + {3 \over 2} \, (\p_1 \cdot \vec n)^2 p_2^2 + {3 \over 2} \, (\p_2 \cdot \vec n)^2 p_1^2 - (\p_1 \cdot \p_2) (\p_1 \cdot \vec n) (\p_2 \cdot \vec n) \right ]~; \nonumber \\
\label{cor3}
&& \sum_{\rm pol}\,\left | A[e^+e^- \to \Upsilon(5S) \to ^3\!D_3 \, \pi \pi] \right |^2= \\ \nonumber 
&& {2  C \over 15} \,\left [ 16 p_1^2 p_2^2 + 4 \, (\p_1 \cdot \p_2)^2 - 3 \,(\p_1 \cdot \vec n)^2 p_2^2 - 3 \,(\p_2 \cdot \vec n)^2 p_1^2 + 2  \, (\p_1 \cdot \p_2) (\p_1 \cdot \vec n) (\p_2 \cdot \vec n) \right ]~, 
\eea
where the sum runs over the polarizations of the final $^3\!D_J$ resonances and the constant $C$ is the same in these expressions, so that the total transition rate to a state with fixed $J$ is proportional to the statistical weight $2J+1$. By averaging over the directions of one pion momentun, these triple correlations can be readily simplified to expressions for the angular distribution over the angle $\theta$ between  a single pion momentum and the direction of the beams:
\bea
&&  {d \over d \cos \theta} \Gamma \left [e^+e^- \to \Upsilon(5S) \to ^3\!D_1 \, \pi \pi \right] \, \propto  \,   1 - {21 \over 47 } \, \cos^2 \theta ~; \nonumber \\
&& {d \over d \cos \theta} \Gamma \left [e^+e^- \to \Upsilon(5S) \to ^3\!D_2 \, \pi \pi \right] \, \propto  \,  1 + {7 \over 11 } \, \cos^2 \theta ~; \nonumber \\
&&  {d \over d \cos \theta} \Gamma \left [e^+e^- \to \Upsilon(5S) \to ^3\!D_3 \, \pi \pi \right] \, \propto  \,  1 - {1 \over 7 } \, \cos^2 \theta ~,
\label{cor1}
\eea

\end{document}